\documentclass{PoS}
\pdfoutput=1

\title{\vspace{-3.5cm}{\normalsize SLAC--PUB--14319}\\\vspace{2cm}
Multi-jet merging with NLO matrix elements}

\ShortTitle{Multi-jet merging with NLO matrix elements}

\usepackage{xspace}
\usepackage{mciteplus}
\usepackage{amsmath}
\usepackage{amssymb}
\newcommand{\POWHEG}{P\protect\scalebox{0.8}{OWHEG}\xspace}
\newcommand{\Sherpa}{S\protect\scalebox{0.8}{HERPA}\xspace}
\newcommand{\MENLOPS}{ME\protect\scalebox{0.8}{NLO}PS\xspace}

\author{\speaker{Frank Siegert}\thanks{Supported by the MCnet Marie Curie Research Training Network MRTN-CT-2006-035606.}\\
        Physikalisches Institut, Albert-Ludwigs-Universit{\"a}t Freiburg\\
        E-mail: \email{frank.siegert@cern.ch}}

\author{Stefan H{\"o}che\\
        SLAC National Accelerator Laboratory, Stanford University\\
        E-mail: \email{shoeche@slac.stanford.edu}}

\author{Frank Krauss\\
        IPPP, Durham University\\
        E-mail: \email{frank.krauss@durham.ac.uk}}

\author{Marek Sch{\"o}nherr\\
        Institut f{\"u}r Kern- und Teilchenphysik, Technische Universit{\"a}t Dresden\\
        E-mail: \email{marek.schoenherr@tu-dresden.de}}

\abstract{In the algorithm presented here, the
  ME+PS approach~\cite{Hoeche:2009rj} to merge samples of tree-level
  matrix elements into inclusive event samples is combined with the \POWHEG
  method~\cite{Nason:2004rx}, which includes exact
  next-to-leading order matrix elements in the parton shower.
  The advantages of the method are discussed and the quality of its
  implementation in \Sherpa is exemplified by results for $e^+e^-$ annihilation
  into hadrons at LEP, for deep-inelastic lepton-nucleon scattering at HERA,
  for Drell-Yan lepton-pair production at the Tevatron
  and for $W^+W^-$-production at LHC energies.
}

\FullConference{35th International Conference of High Energy Physics\\
                 July 22-28, 2010\\
                 Paris, France}

\begin{document}

\section{Introduction}

The simulation of hard QCD radiation in parton-shower Monte Carlos has
seen tremendous progress over the last years. It was largely stimulated by the
need for more precise predictions at LHC energies where the large available
phase space allows additional hard QCD radiation alongside known Standard Model
processes or even signals from new physics.

Two types of algorithms have been developed, which allow to improve upon the
soft-collinear approximations made in the parton shower, such that hard radiation
is simulated according to exact matrix elements.
In the ME+PS
approach~\cite{Hoeche:2009rj} higher-order tree-level matrix elements for
different final-state jet multiplicity are merged with each other and with
subsequent parton shower emissions to generate an inclusive sample. Such a
prescription is invaluable for analyses which are sensitive to final states with
a large jet multiplicity. The only remaining deficiency of such tree-level
calculations is the large uncertainty stemming from scale variations.
The \POWHEG method~\cite{Nason:2004rx,*Frixione:2007vw} solves this problem
for the lowest multiplicity subprocess
by combining full NLO matrix elements with the parton shower. While this leads
to NLO accuracy in the inclusive cross section and the exact radiation pattern
for the first emission, it fails to describe higher-order emissions with improved
accuracy. Thus it is not sufficient if final states with high jet
multiplicities are considered.

With the complementary advantages of these two approaches, the question arises
naturally whether it would be possible to combine them into an even more
powerful one. Such a combined algorithm was independently developed
in~\cite{Hamilton:2010wh} and~\cite{Hoeche:2010kg}. Here a summary of the
algorithm is given and predictions from corresponding Monte-Carlo predictions
are presented.

\section{The \MENLOPS formalism}
 
To combine both, the ME+PS and POWHEG method, one has to work out how they contribute to a cross
section including up to the first emission off a given core process.
All further emissions are not affected by the \POWHEG approach, and can thus be
treated as before in ME+PS.

The master formula for the expectation value of an observable in the \POWHEG
method contains two terms contributing to the cross section at NLO and can be
written down in a simplified form (for details the reader is referred
to~\cite{Nason:2004rx,*Frixione:2007vw},~\cite{Hoeche:2010pf}):
\begin{equation}\label{Eq:master_powheg}
  \langle O\rangle^{\rm POW}=\sum_i\int{\rm d}\Phi_B\,\bar{\rm B}_i(\Phi_B)
  \Bigg[\; 
    \underbrace{\bar{\Delta}_i(t_0)\,O(\Phi_B)}_\textrm{no emission} + 
    \underbrace{\sum_j\int_{t_0} {\rm d}\Phi_{R|B}\,\frac{{\rm R}_j(\Phi_R)}{{\rm B}_i(\Phi_B)}\;
      \bar{\Delta}_i(t)\,O(\Phi_R)\;}_\textrm{resolved emission}
  \Bigg]\;,
\end{equation}
where $\bar{\rm B}_i(\Phi_B)$ is the differential cross section at NLO for the 
Born phase-space configuration $\Phi_B$ and 
$\bar{\Delta}_i(t)=\exp\left\{\,-\sum_j\int_t{\rm d}\Phi_{R|B}\,\rm R_j/B_i\,\right\}$ 
is the matrix-element-corrected Sudakov form factor. The indices $i$ and $j$ label parton
configurations, see~\cite{Hoeche:2010pf}. The radiative phase space $\Phi_{R|B}$ is parametrised
by variables in a parton-shower picture, including the ordering variable $t$ which is bounded from
below by the cut-off $t_0\sim\Lambda_\text{QCD}^2$, regularising the integral.

The same quantity can be calculated in the ME+PS approach again only taking into
account the first emission:
\begin{equation}\label{Eq:master_meps}
  \begin{split}
  &\langle O\rangle^{\rm ME+PS}=\sum_i\int{\rm d}\Phi_B\,{\rm B}_i(\Phi_B)
  \Bigg[\; 
    \underbrace{\Delta_i(t_0)\,O(\Phi_B)}_\text{no emission} + 
    \sum_j\int_{t_0}{\rm d}\Phi_{R|B}\\
    &\qquad\times\Big(
      \underbrace{\Theta(Q_{\rm cut}-Q)\,\frac{8\pi\alpha_s}{t}\,
        \mathcal{K}_{R_j|B_i}\,\frac{\mathcal{L}_{R_j}}{\mathcal{L}_{B_i}}}_{\textrm{PS domain}}+
      \underbrace{\Theta(Q-Q_{\rm cut})\,
        \frac{{\rm R}_j(\Phi_R)}{{\rm B}_i(\Phi_B)}}_{\text{ME domain}}
    \Big)\,\Delta_i(t)\,O(\Phi_R)
  \;\Bigg]\;.
  \end{split}
\end{equation}
The term for resolved emissions is now separated into the ``ME domain'' and ``PS domain''
and describes the probability of additional 
QCD radiation according to the real-radiation matrix elements and their corresponding 
parton-shower approximations, respectively. $\mathcal{K}_{R_j|B_i}$ denote
the evolution kernels of the parton shower and $\mathcal{L}_{R_j/B_i}$ are the parton luminosities 
of the real-emission and the underlying Born configurations. The Sudakov form factor
$\Delta_i(t)$ is given by the parton shower and does not include matrix-element
corrections, in contrast to $\bar{\Delta}_i(t)$ in Eq.~\eqref{Eq:master_powheg}.

The MENLOPS approach then combines the two above equations:
\begin{equation}\label{Eq:master_menlops}
  \begin{split}
  \langle O\rangle^{\rm MENLOPS}=&\sum_i\int{\rm d}\Phi_B\,\bar{\rm B}_i(\Phi_B)
  \Bigg[\; 
    \underbrace{\bar{\Delta}_i(t_0)\,O(\Phi_B)}_\text{no emission} + 
    \sum_j\int_{t_0}{\rm d}\Phi_{R|B}\,\frac{{\rm R}_j(\Phi_R)}{{\rm B}_i(\Phi_B)}\\
    &\qquad\times\Big(
      \underbrace{\Theta(Q_{\rm cut}-Q)\,\bar{\Delta}_i(t)}_{\text{PS domain}}+
      \underbrace{\Theta(Q-Q_{\rm cut})\,\Delta_i(t)}_{\text{ME domain}}
    \Big)\,O(\Phi_R)
  \;\Bigg]\;.
  \end{split}
\end{equation}
The cross section of such a sample is determined by the NLO weight $\bar{B}$ as
in \POWHEG and could only be modified due to the appearance of $\Delta_i(t)$
instead of $\bar{\Delta}_i(t)$ in the ``ME domain''. Expanding the induced
correction factor
$\bar{\Delta}_i(t)/\Delta_i(t)$ to first order reveals that Eq.~\eqref{Eq:master_menlops}
automatically yields next-to-leading order accurate predictions for any infrared
and collinear safe observable $O$. At the same time it becomes straightforward
to include higher-order tree-level matrix elements to improve the description of
high-multiplicity jet final states.

\section{Results}

\begin{figure}
  \centering
  \includegraphics[width=0.4\textwidth]{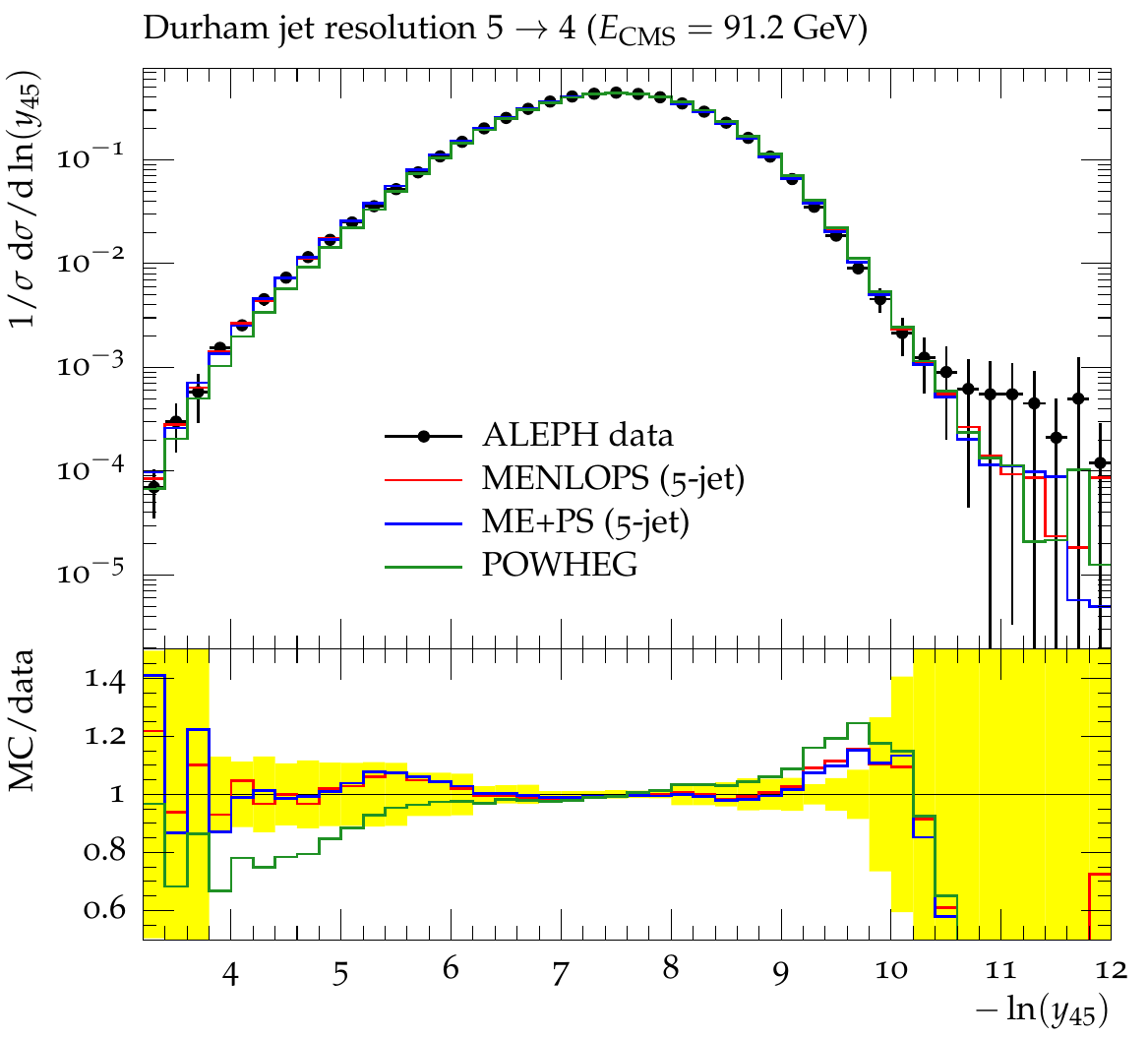}
  \includegraphics[width=0.4\textwidth]{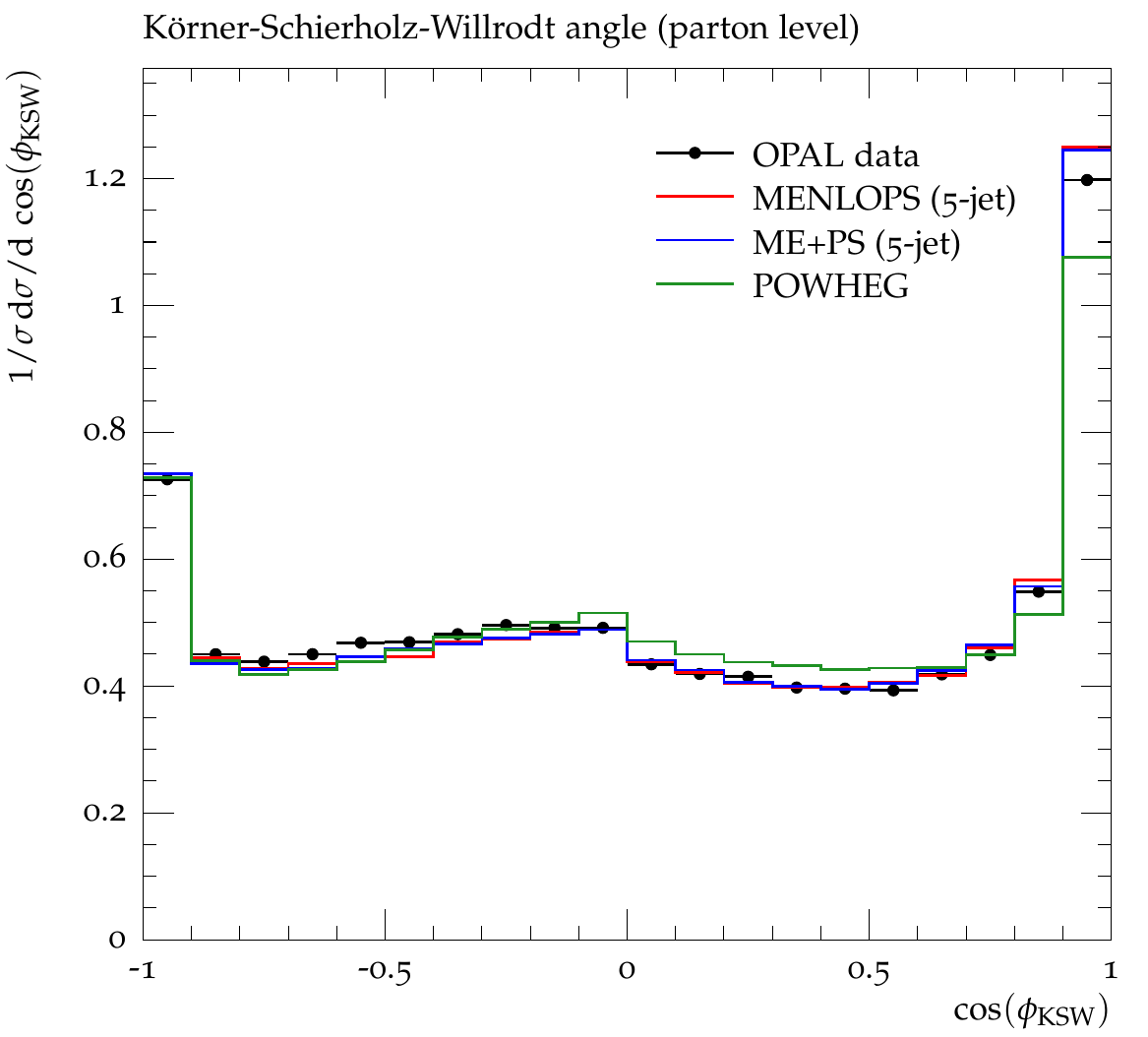}
  \caption{Jet resolution for $5\to 4$ clusterings in the Durham algorithm (left) and KSW angle between the four hardest jets (right) in $e^+e^-\to$ hadrons at $\sqrt{s}=$91.25 GeV.}
  \label{fig:ee}
\end{figure}

\begin{figure}
  \centering
  \includegraphics[width=0.4\textwidth]{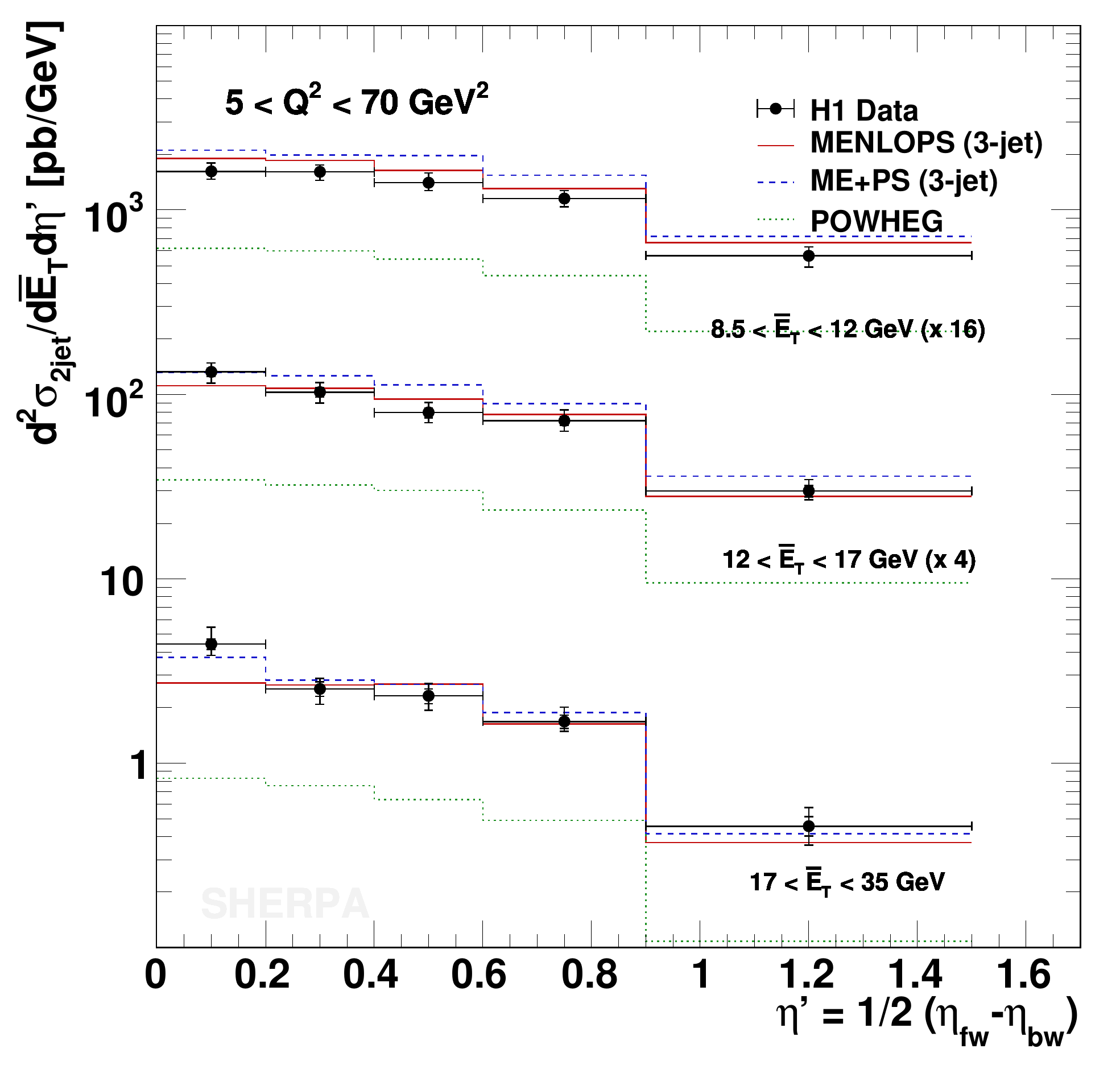}\nolinebreak
  \includegraphics[width=0.4\textwidth]{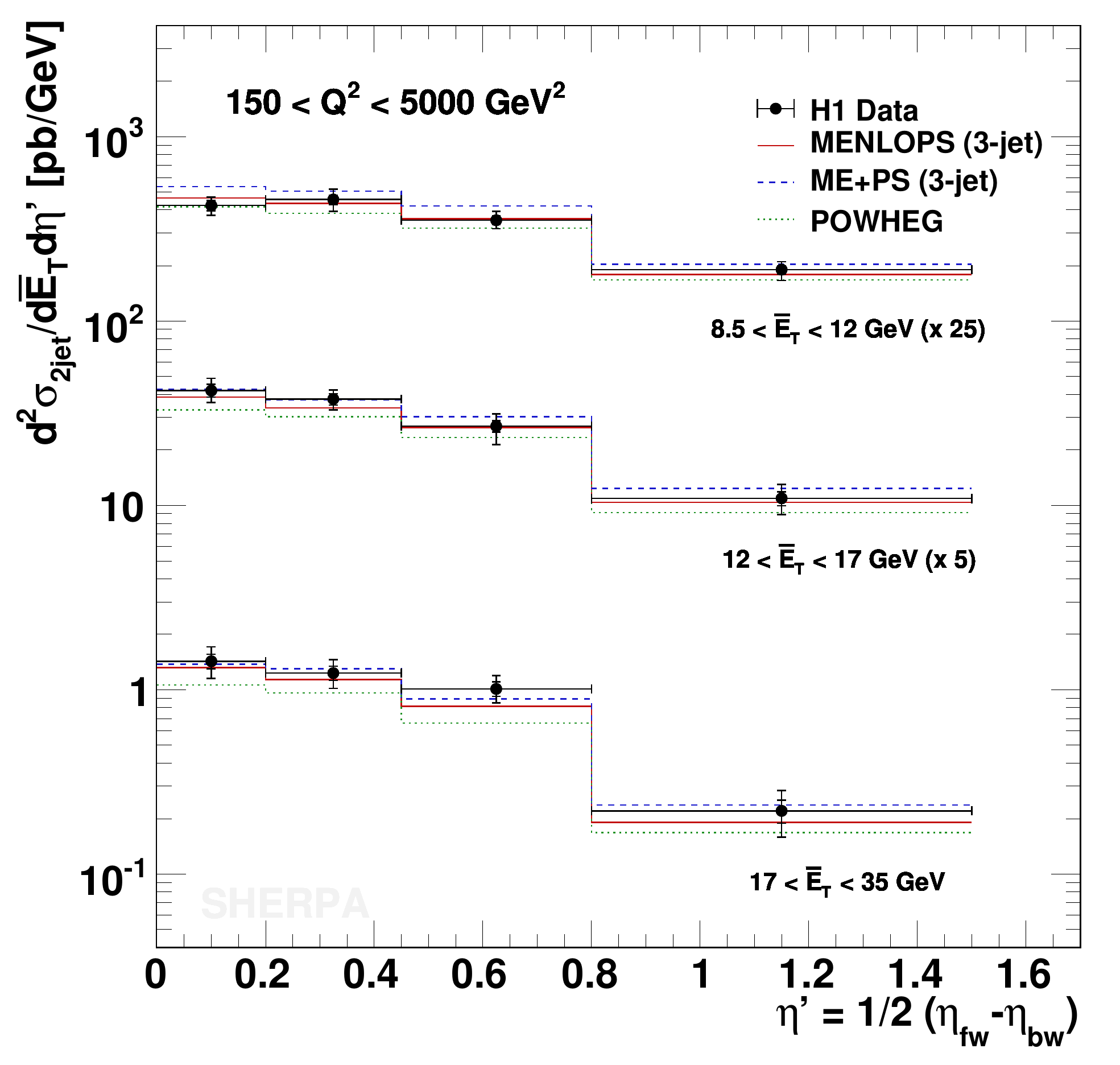}
  \caption{Pseudorapidity difference between the forward and backward jet in
    DIS dijet events for $e^+p\to e^+$+jets at $\sqrt{s}=$~300~GeV.}
  \label{fig:dis}
\end{figure}

\begin{figure}
  \centering
  \includegraphics[width=0.35\textwidth]{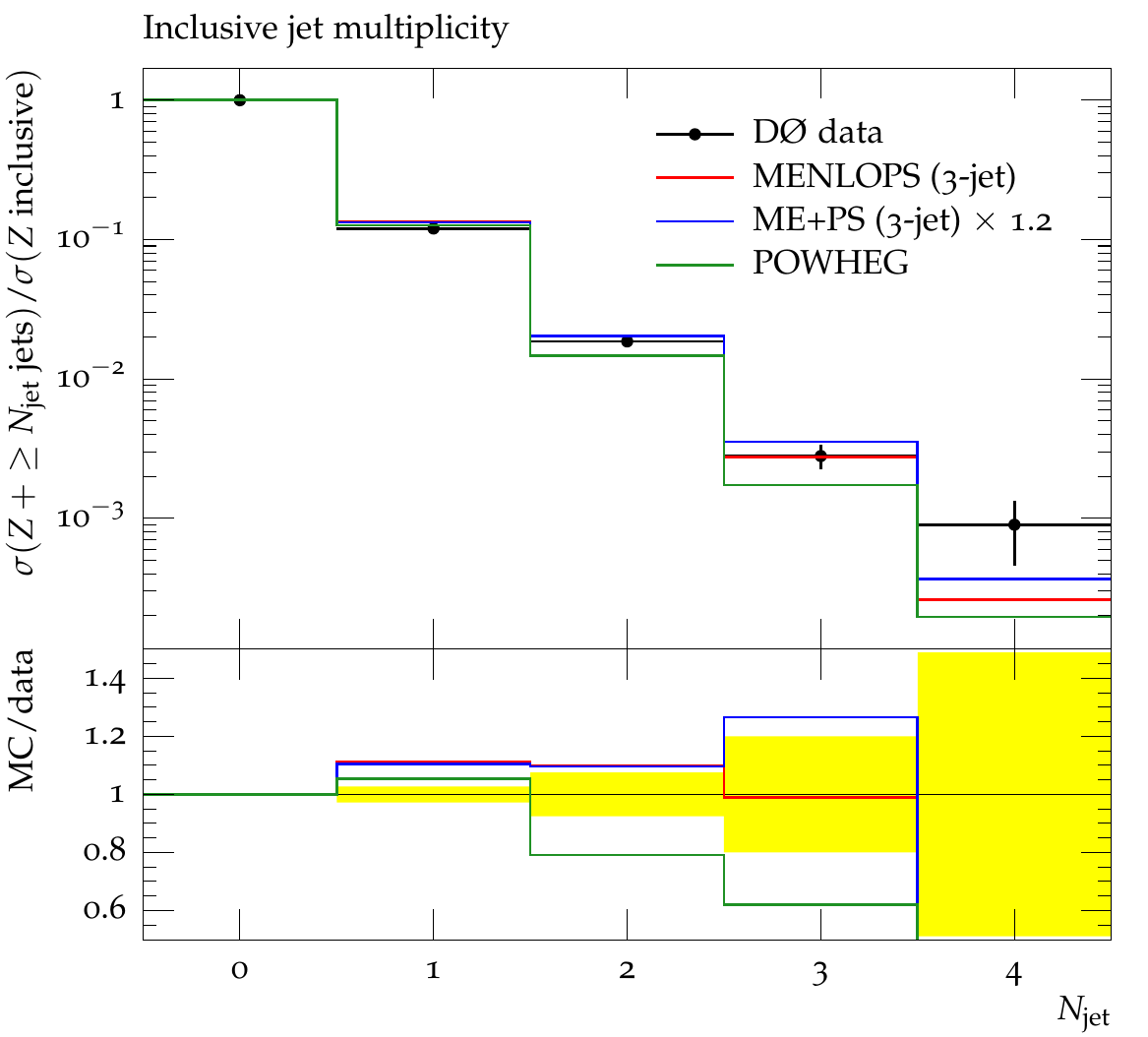}
  \includegraphics[width=0.35\textwidth]{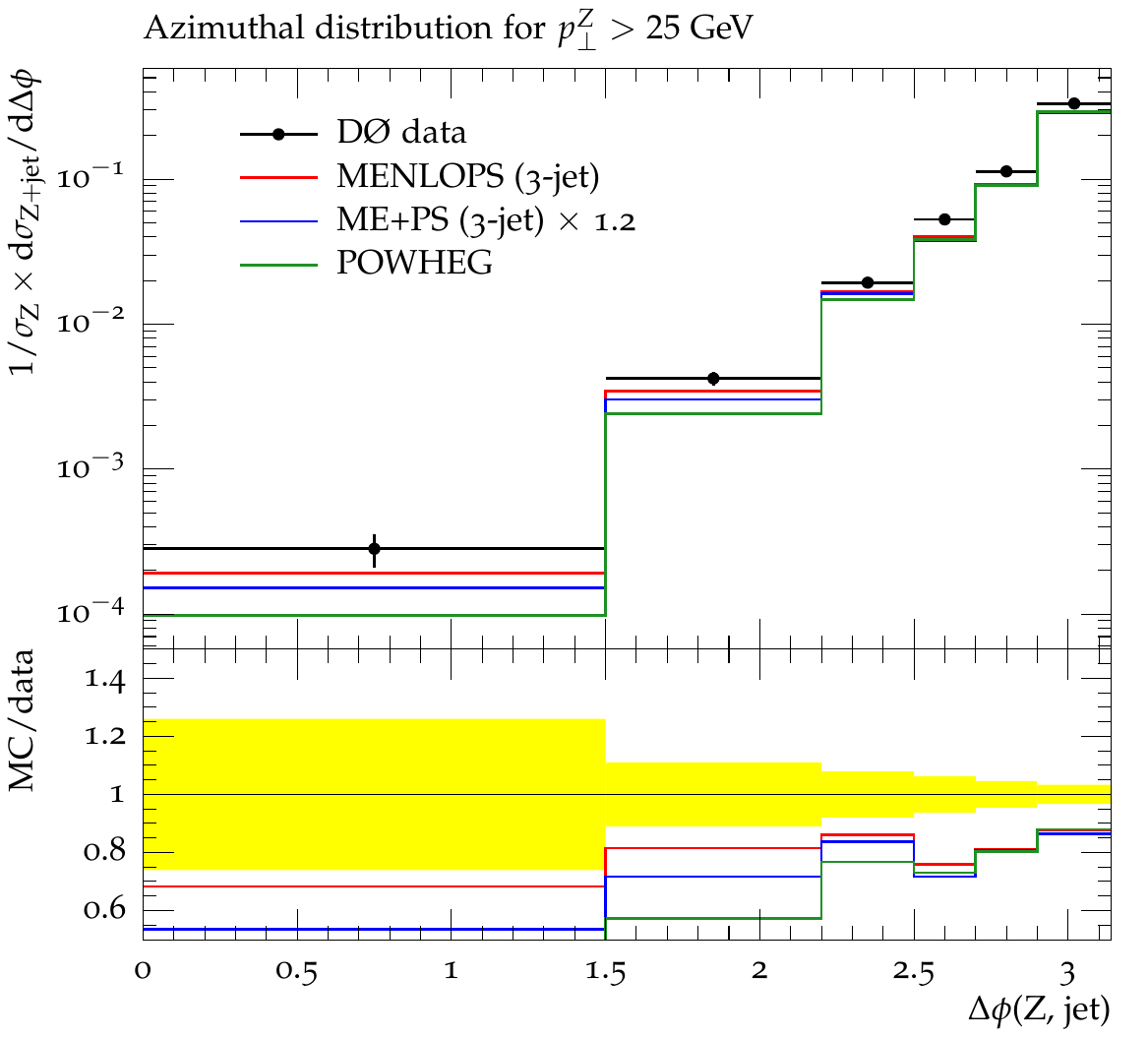}
  \caption{Inclusive jet multiplicity (left) and azimuthal separation of lepton pair and leading jet (right) in $\bar{p}p\to \ell\ell$+jets at $\sqrt{s}=$1.96 TeV}
  \label{fig:zjets}
\end{figure}

\begin{figure}
  \centering
  \includegraphics[width=0.35\textwidth]{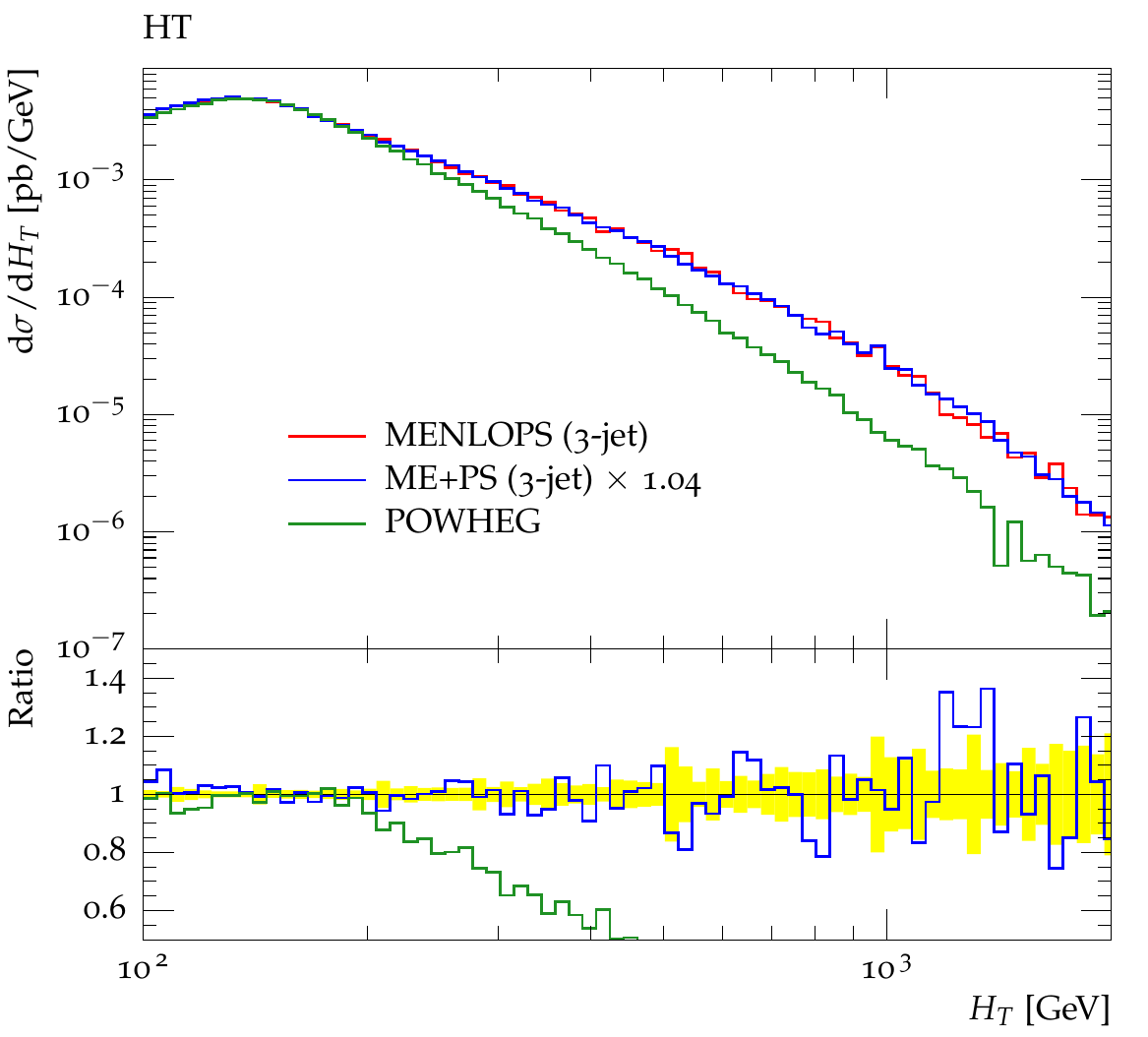}
  \includegraphics[width=0.35\textwidth]{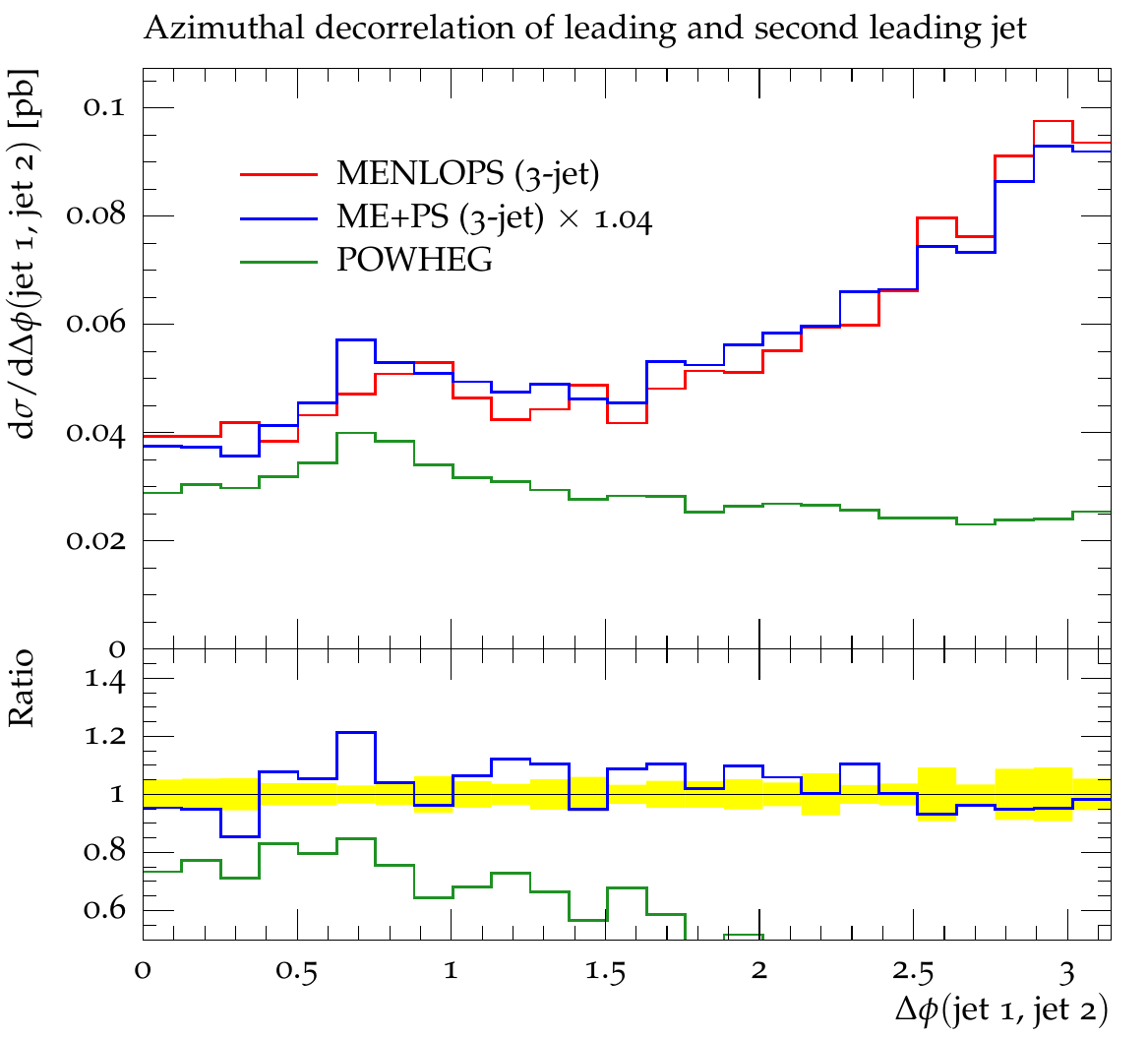}
  \caption{Scalar sum of missing $E_T$ and transverse momenta of jets and leptons (left) and azimuthal decorrelation between leading and second leading jet (right) in $pp\to W^+W^-$+jets at $\sqrt{s}=$14 TeV}
  \label{fig:ww}
\end{figure}

The following collection of results showcases the improved description of
experimental data as well as the significant difference to pure \POWHEG
samples due to the improved description of additional hard radiation.
We employ the Monte-Carlo event generator \Sherpa~\cite{Gleisberg:2003xi,*Gleisberg:2008ta},
which provides
automated implementations of both ME+PS merging and the \POWHEG
algorithm as reformulated in~\cite{Hoeche:2010pf}.
Fig.~\ref{fig:ee} displays results for electron-positron annihilation into
hadrons at $\sqrt{s}=$91.25 GeV. In both, the jet resolution distribution for
$5\to 4$ clusterings in the Durham algorithm and the KSW angle between the
four hardest jets, the improved description from exact matrix elements including
all correlations is exemplified.
Results for deep-inelastic lepton-nucleon
scattering (DIS) are presented in Fig.~\ref{fig:dis}. We show the pseudorapidity difference
between the forward and backward jet in dijet events. Since these distributions are differential in
the jet transverse energy in the Breit frame, they provide an excellent test of the correct
simulation of jet activity in the Monte-Carlo. We observe a drastic difference between the
\POWHEG and MENLOPS predictions. Only the improved simulation of additional
hard QCD radiation through the MENLOPS approach allows to achieve agreement with experimental data.
For jet production in association with a
Drell-Yan lepton pair at Tevatron Run 2 energies one can find similar results in
Fig.~\ref{fig:zjets}. Both, the inclusive jet multiplicity as well as the azimuthal
separation of the lepton pair and the leading jet are described significantly
better when higher-multiplicity matrix elements are included. Turning to
predictions for the LHC, Fig.~\ref{fig:ww} shows the scalar sum of missing $E_T$
and transverse momenta of jets and leptons as well as the azimuthal
decorrelation between the leading and second leading jet in $W^+W^-$ production.
Both observables are
significantly influenced by final states with many jets, and thus show a strong
difference between the \POWHEG and MENLOPS results.
For all processes investigated
here, the results of the tree-level ME+PS approach resemble the features of the
MENLOPS results as long as they are multiplied with a global $K$-factor as
indicated in the plot legends.

\bibliographystyle{bib/amsunsrt_modp}  
\bibliography{bib/journal}

\ifx\mcitethebibliography\mciteundefinedmacro
\PackageError{amsunsrt_mod.bst}{mciteplus.sty has not been loaded}
{This bibstyle requires the use of the mciteplus package.}\fi
\begin{mcitethebibliography}{1}

\bibitem{Hoeche:2009rj}
S.~H{\"o}che, F.~Krauss, S.~Schumann and F.~Siegert, \emph{{QCD matrix elements
  and truncated showers}}, JHEP \textbf{05} (2009),
  \href{http://www.slac.stanford.edu/spires/find/hep/www?eprint=arXiv:0903.121%
9}{053},  [\href{http://arXiv.org/pdf/0903.1219}{{\tt arXiv:0903.1219}}
  [hep-ph]]%
\relax\mciteBstWouldAddEndPuncttrue
\mciteSetBstMidEndSepPunct{\mcitedefaultmidpunct}
{\mcitedefaultendpunct}{\mcitedefaultseppunct}\relax
\EndOfBibitem
\bibitem{Nason:2004rx}
P.~Nason, \emph{{A new method for combining NLO QCD with shower Monte Carlo
  algorithms}}, JHEP \textbf{11} (2004),
  \href{http://www.slac.stanford.edu/spires/find/hep/www?eprint=hep-ph/0409146%
}{040},  [\href{http://arXiv.org/pdf/hep-ph/0409146}{{\tt hep-ph/0409146}}]%
\relax\mciteBstWouldAddEndPuncttrue
\mciteSetBstMidEndSepPunct{\mcitedefaultmidpunct}
{\mcitedefaultendpunct}{\mcitedefaultseppunct}\relax
\EndOfBibitem
\bibitem{Frixione:2007vw}
S.~Frixione, P.~Nason and C.~Oleari, \emph{{Matching NLO QCD computations with
  parton shower simulations: the POWHEG method}}, JHEP \textbf{11} (2007),
  \href{http://www.slac.stanford.edu/spires/find/hep/www?eprint=arXiv:0709.209%
2}{070},  [\href{http://arXiv.org/pdf/0709.2092}{{\tt arXiv:0709.2092}}
  [hep-ph]]%
\relax\mciteBstWouldAddEndPuncttrue
\mciteSetBstMidEndSepPunct{\mcitedefaultmidpunct}
{\mcitedefaultendpunct}{\mcitedefaultseppunct}\relax
\EndOfBibitem
\bibitem{Gleisberg:2003xi}
T.~Gleisberg, S.~H{\"o}che, F.~Krauss, A.~Sch{\"a}licke, S.~Schumann and
  J.~Winter, \emph{{\Sherpa 1.$\alpha$, a proof-of-concept version}}, JHEP
  \textbf{02} (2004),
  \href{http://www.slac.stanford.edu/spires/find/hep/www?irn=5730570}{056},
  [\href{http://arXiv.org/pdf/hep-ph/0311263}{{\tt hep-ph/0311263}}]%
\relax\mciteBstWouldAddEndPuncttrue
\mciteSetBstMidEndSepPunct{\mcitedefaultmidpunct}
{\mcitedefaultendpunct}{\mcitedefaultseppunct}\relax
\EndOfBibitem
\bibitem{Gleisberg:2008ta}
T.~Gleisberg, S.~H{\"o}che, F.~Krauss, M.~Sch\"{o}nherr, S.~Schumann,
  F.~Siegert and J.~Winter, \emph{{Event generation with \Sherpa 1.1}}, JHEP
  \textbf{02} (2009),
  \href{http://www.slac.stanford.edu/spires/find/hep/www?eprint=0811.4622}{007%
},  [\href{http://arXiv.org/pdf/0811.4622}{{\tt arXiv:0811.4622}} [hep-ph]]%
\relax\mciteBstWouldAddEndPuncttrue
\mciteSetBstMidEndSepPunct{\mcitedefaultmidpunct}
{\mcitedefaultendpunct}{\mcitedefaultseppunct}\relax
\EndOfBibitem
\bibitem{Hoeche:2010pf}
S.~H{%
\"o}che, F.~Krauss, M.~Sch{\"o}nherr and F.~Siegert, \emph{{Automating the
  \POWHEG method in \Sherpa}},  \href{http://arXiv.org/pdf/1008.5399}{{\tt
  arXiv:1008.5399}} [hep-ph]%
\relax\mciteBstWouldAddEndPuncttrue
\mciteSetBstMidEndSepPunct{\mcitedefaultmidpunct}
{\mcitedefaultendpunct}{\mcitedefaultseppunct}\relax
\EndOfBibitem
\bibitem{Hamilton:2010wh}
K.~Hamilton and P.~Nason, \emph{{Improving NLO-parton shower matched
  simulations with higher order matrix elements}}, JHEP \textbf{06} (2010),
  \href{http://www.slac.stanford.edu/spires/find/hep/www?eprint=arXiv:1004.176%
4}{039},  [\href{http://arXiv.org/pdf/1004.1764}{{\tt arXiv:1004.1764}}
  [hep-ph]]%
\relax\mciteBstWouldAddEndPuncttrue
\mciteSetBstMidEndSepPunct{\mcitedefaultmidpunct}
{\mcitedefaultendpunct}{\mcitedefaultseppunct}\relax
\EndOfBibitem
\bibitem{Hoeche:2010kg}
S.~H{%
\"o}che, F.~Krauss, M.~Sch{\"o}nherr and F.~Siegert, \emph{{NLO matrix
  elements and truncated showers}},  \href{http://arXiv.org/pdf/1009.1127}{{\tt
  arXiv:1009.1127}} [hep-ph]%
\relax\mciteBstWouldAddEndPuncttrue
\mciteSetBstMidEndSepPunct{\mcitedefaultmidpunct}
{\mcitedefaultendpunct}{\mcitedefaultseppunct}\relax
\EndOfBibitem
\end{mcitethebibliography}

\end{document}